\documentstyle[12pt,a4]{article}

\begin{document}

\begin{center}
{\large \bf The effective action of $W_3$-gravity.}
\vspace{2mm}

{\bf D.R.Karakhanyan}\\

{\small {\em Yerevan Physics Institute, Republic of Armenia \\
(Alikhanian Brothers St. 2,  Yerevan 375036,  Armenia)}\\
E-mail: karakhian@vx1.yerphi.am}
\end{center}

\begin {abstract}
A new method for integrating anomalous Ward identities 
and finding the effective action is proposed. 
Two-dimensional supergravity and $W_3$-gravity are used as 
examples to demonstrate its potential. An operator is 
introduced that associates each physical quantity with a
Ward identity, i.e., a quantity that is transformed without
anomalous terms and can be nullified in a consistent manner.
A covariant form of the action for matter field interacting
with a gravitational and $W_3$-gravitational background is
proposed.
\end{abstract}

\section{INTRODUCTION}

The tremendous upsurge of interest in $W_N$-algebras \cite{z},
that followed their discovery by
A.B.Zamolodchikov can be explained by the fact that the basic
relationships in $W_N$-algebras, in contrast to those of ordinary 
Lie algebras, are multilinear and that the mathematical aspects 
had not been systematically studied. A big achievment in this area 
of research was the use of the Drinfel'd-Sokolov reduction scheme
\cite{GD}, which reduces $W$-algebras to Lie algebras and relates
them to the second hamiltonian structure of the generalized
Korteweg-de Vries hierarchies; $W_N$-algebras contain Virasoro 
algebra as a subalgebra. In the context of string theory, the 
appearence of the latter is a reflection of invariance under 
reparametrization of the string world surface. The extension of 
this symmety to invariance under $W$-gravity transformations leads to 
the theory of $W$-strings in Polyakov approach, i.e., to the theory of 
interaction of matter fields with ordinary (spin-2) and $W$-gravitational 
(spin-$N$) background fields. Thus, symmetry under transformations of
$W$-gravity is the leading principle that makes it possible to write the 
interaction for the fields with spin $\geq 2$, at least in two dimensions.

However, progress in this area of research was fraught with considerable
difficulties. First the chiral theory of the interaction of matter  
and $W$-gravity was formulated by Hull \cite{h}. Then Schoutens 
et al. \cite{ssn1} generalized the theory  to the non-chiral  case 
but encounted significant technical difficulties: the action in the 
theory proved to be infinitely nonlinear in the matter fields and 
non-local, so any futher  analysis is extremely complicated.
By calculculating the functional integral over the matter fieldsI with a 
central charge $c$ interacting with $W_3$-gravity Schotens et. al.
also found the induced action of $W_3$-gravity in the form of a 
$1/c$ expansion \cite{ssn2}. The same researchers (see Ref. [6]) 
found the induced action of chiral $W_3$-gravity, a direct analog of the 
Polyakov's action \cite{pol} for ordinary gravity, by integrating the 
anomaly in the limit $c\rightarrow\infty$.

Clarification of the geometrical meaning of $W$-tranformations would help 
$W$-gravity  studes considerably. This aspect was studied by 
Figueroa-O'Farrill et al. and Hull \cite{fh}.

At present it is generally hoped  that $W$-gravity studies will help to 
overcome  the strong coupling barrier $c=1$ for a system consisting of 
conformal matter and two-dimensional gravitation, which will probably 
make it possible to avoid the fractional dimensionality 
estabilished by Knizhnik and et. al. \cite{kpz} for quantum gravity
in in the weak-coupling mode. Direct generalization of the results of 
Ref. [9] to $W$-gravity in the absence of matter fields done by Matsuo \cite{m}.

The  present investigation develops a method for integrating two-dimensional 
anomalous Ward identities. Its aplication is illustrated by examples of 
two-dimensional gravity, supegravity, and $W_3$-gravity. The essence of the 
method consists in the following . By expressing anomalous currents in terms 
of free fields  via bosonization formulae, we can lower the order of these 
differnital equations and integrate them. The resulting effective action
reproduces the anomaly correctly. When the regularization scheme changes, 
local counterterms are added to the non-local effective action, and the
emergence of these counterterms changes the form and symmetry of the Ward
identities. The bosonized fields, being free in one regularization scheme, 
in another scheme are related by the fact that they satisfy certain Ward 
identities. When the chiral Weyl-invariant regularization scheme is replaced 
by the diffeomorphysm-invariant scheme, local counterterms are added in such 
a way that the kinetic part of the effective action becomes invariant both
under diffeomorphysm and under Weyl transformations. The remaining (topological) 
part of the effective action is fixed by requirement that the total action , 
being diffeomorphysm-invariant, under Weyl transformations, be symmetric in 
the quantum or projective sence, i.e., is transformed as a 1-cocycle.

In Sect. 2 and 3 the application of this method is demonstraited using the
well-known examples of ordinary and (N=1)-supergravity, and a differential
operator $R$ is introduced, which with each physical quantity assosiates its
Ward identity. The operator is actually a Slavnov operator, which was studied
by Zucchini \cite{zu} in connection with two-dimensional gravity in
cojunction with an additional inhomogeneous term that destroys the 
anomalous contribution in the transformation law.

In Sect. 4 these calculations are generalized to the case of chiral 
$W_3$-gravity. It was found that the result is in full agreement with
that of Ooguri et. al. \cite{ossn}.

Finally, in Sect. 5 deals with the covariant action of matter interacting 
with nonchiral $W_3$-gravity. In addition to exibiting parametrization
symmetry and $W$-diffeomorphysm symmetry, this action is $W$-Weyl invariant
and can serve as the kinetic part of the effective action of $W_3$-gravity 
calculated in the diffeomorphysm-invariant scheme.
\section{Two-dimensional gravity}

The Polyakov action, which was derived in Ref. [7] as the effective action 
induced by chiral matter interacting with two-dimensional gravity, is the 
determinant of the two-dimensional Laplasian calculated in a regularization
scheme that conserves Weyl symmetry and half the reparametrization symmetry.
The presence of a conformal anomaly manifests induces an explicit dependence
of the Polyakov action on one of the reparametrization functions. In other 
words, this effective action can be calculated by integrating the apprppiate 
variational equation, the Ward identity.

The Ward identity of two-dimensional gravitation theory in the light-cone 
gauge is well known:
\begin{equation}
\label{1}
R_T=(\bar\partial -h\partial -2\partial h)T-\partial^3h=0.
\end{equation}
It expresses the anomalous conservation of the system's energy-momentum tensor
$T$. The field $h$ in this expression denotes the nonvanishing metric component 
that remains after light-cone gauge is specified. It is covariant under the
transformations
\begin{eqnarray}
\label{2}
&&\delta h=(\bar\partial-h\partial+\partial h)\epsilon,\nonumber\\
&&\delta T=(\partial^3+2T\partial+\partial T)\epsilon,
\end{eqnarray}
i.e
\begin{equation}
\label{3}
\delta_\epsilon R_T=(\epsilon\partial+2\partial\epsilon)R_T.
\end{equation}
Equation (\ref{3}) expresses the Wess-Zumino self-consistency condition.
If we use the bosonization formula and parametrize the energy-momentum tensor
via a scalar field,
\begin{equation}
\label{4}
T=\partial^2\varphi-\frac{1}{2}(\partial\varphi)^2,
\end{equation}
the order of the anomalous term in (\ref{1}) can be reduced:
\begin{eqnarray}
\label{5}
&R_\varphi=(\bar\partial -h\partial)\varphi -\partial h=0,&\\
&\delta_\epsilon R_\varphi=\epsilon\partial R_\varphi.& \nonumber
\end{eqnarray}
Comparing (\ref{1}) with (\ref{4}) and (\ref{5}), we obtain
\begin{equation}
\label{6}
R_T=\partial^2R_\varphi-\partial\varphi\partial R_\varphi.
\end{equation}
The transformation law for the scalar field $\varphi$ is also anomalous:
\begin{equation}
\label{7}
\delta\varphi=\partial\epsilon+\epsilon\partial\varphi.
\end{equation}
If for the field $\varphi$ we postulate the free-field Poisson bracket,
\begin{equation}
\label{8}
\{\partial\varphi(x);\partial\varphi(x^\prime
)\}=\delta^\prime(x-x^\prime),
\end{equation}
the $\epsilon$-variation of any quantity $A$ can be determined by its Poisson
bracket with energy-momentum tensor:
\begin{equation}
\label{9}
\delta_\epsilon A(ì)=\int d^2x^\prime\epsilon(x^\prime).
\{T(x^\prime);A(x)\}
\end{equation}
Clearly, this definition for the field $\varphi$ coincides with (\ref{7}).
The bracket of tensor $T$ with itself is
\begin{equation}
\label{10}
-\{T(x);T(x^\prime)\}=\delta^{\prime\prime\prime}(x-x^\prime)+
(T(x)+T(x^\prime))\delta^\prime(x-x^\prime).
\end{equation}
Although the energy-momentum tensor can be expressed in terms of $\varphi$,
there is no way in which we can express the gauge field $h$ in terms 
$\varphi $ in (\ref{5}) in a local manner. To do this we must introduce a 
quantity that satisfies the regular Ward identity, i.e., a quantity that 
transforms as a scalar. The anomaly can be removed from the Ward identity by
introducing a scalar field $f$ in the following way:
\begin{equation}
\label{11}
\varphi=\log\partial f.
\end{equation}
The transformation law for $f$ and the corresponding Ward identity have the 
form
\begin{eqnarray}
\label{12}
&\delta_\epsilon f=\epsilon\partial f,&\nonumber\\
&R_f=(\bar\partial-h\partial)f=0,& \\
&\delta_\epsilon R_f=\epsilon\partial R_f.&\nonumber
\end{eqnarray}
Now, when all of the quantities are expressed in terms of the function $f$ 
locally, we can integrate the variational equation for the effective action of 
the theory, which can also be expressed in terms of $f$ locally and is given 
by the Polyakov formula \cite{pol}. Detailed calculations are given in Sect. 3
for the more interesting case of supergravity.

Comparing Eqs. (\ref{3}), (\ref{6}) and (\ref{12}), we can see that the gauge 
variation $\delta$ and the Ward identities $R$ are commutative operations on 
the fields $T$, $\varphi$ and $f$.

The relationship between $R_f$, $R_\varphi$ and $R_T$ is specified by the 
following formulae:
\begin{eqnarray}
\label{13}
&R_\varphi=\frac{\partial R_f}{\partial f},&\nonumber\\
&R_T=(\partial^3+2T\partial+\partial T)(\frac{R_f}{\partial f}).&
\end{eqnarray}
We see that the operator $R$ assosiates with each physical quantity $X$ a
covariant expression $R_X$, its Ward identity, which in view of its covariance
under gauge transformations can be consistently made to vanish. But since the 
theory lacks quantities of the required dimensionality, this expression must to
be set to zero. Comparing (\ref{13}) with (\ref{11}) and (\ref{6}) with 
(\ref{4}), we can see that $R$ obeys the Newton-Leibniz rule. This property
of $R$ makes it possible to write the Ward identities for the correlation 
functions of the fields $T$, $\varphi$ etc. immediately.

If we apply the Legendre transformation
\begin{equation}
\label{14}
Z[h]=\Gamma[h]+\int d^2xTh,
\end{equation}
Eq. (\ref{1}) can be written as
\begin{equation}
\label{15}
(\partial^3+2T\partial+\partial T)\frac{\delta\Gamma}{\delta T}=
-\bar\partial T,
\end{equation}
where the Bol operator \cite{g} on the left-hand side is the covariant form
of $\partial^3$ on an arbitrary Riemann surface, and contains a projective
connection, for which we may take $T$. This notation expresses the covariance
of the Ward identity (\ref{1}) just as the Wess-Zumino self-consistency 
condition (\ref{3}) does.
\setcounter{equation}{0}
\section{Simple supergravity}

In this section we generalize all the ideas of Sect. 2 to the case of simple
supergravity and calculate the effective action of the theory.

Polyakov's result was generalized by Polyakov and Zamolodchikov \cite{pz}
to the case of (1.0)-supergravity. The corresponding generalization of the 
Polyakov action represents the effective action obtained in a regularization 
scheme that conserves the Weyl and super-Weyl symmetries and half the
supercoordinate symmetry \cite{k}. The nontrivial dependence of the action
on the other coordinate functions (odd- and even-parity) is determined by a 
superconformal anomaly.

The Ward identities of two-dimensional supergravity in the light-cone gauge 
can be written as \cite{ossn}
\begin{eqnarray}
\label{2.1}
R_T=(\bar\partial-h\partial-2\partial h)T-(\frac{1}{2}\chi\partial
+\frac{3}{2}\partial\chi)S-\partial^3h=0,\nonumber\\
R_S=(\bar\partial-h\partial-\frac{3}{2}\partial h)S-\frac{1}{2}
\chi T-\partial^2\chi=0.
\end{eqnarray}
They are covariant under the transformations
\begin{eqnarray}
\label{2.2}
&&\delta h=(\bar\partial-h\partial+\partial h)\epsilon+\frac{1}{2}
\kappa\chi,\nonumber\\
&&\delta\chi=(\bar\partial-h\partial+\frac{1}{2}\partial h)\kappa+
(\epsilon\partial-\frac{1}{2}\partial\epsilon)\chi,\nonumber\\
&&\delta T=(\partial^3+2T\partial+\partial T)\epsilon+
(\frac{1}{2}\kappa\partial+\frac{3}{2}\partial \kappa)S,\\
&&\delta S=(\epsilon\partial+\frac{3}{2}\partial\epsilon)S+
(\partial^2+\frac{1}{2}T)\kappa, \nonumber
\end{eqnarray}
i.e.,
\begin{eqnarray}
\label{2.3}
&&\delta R_T=(\epsilon\partial+2\partial\epsilon)R_T+
(\frac{1}{2}k\partial+\frac{3}{2}\partial\kappa)R_S,\nonumber\\
&&\delta R_S=(\epsilon\partial+\frac{3}{2}\partial\epsilon)R_S+
\frac{1}{2}\kappa R_T,
\end{eqnarray}
which means that the Ward identity $R_A$ transforms in the same way as the
quantity $A$ but without anomalous terms.

Going over to the scalar multiplet of matter fields $(\varphi,\lambda)$, with
\begin{eqnarray}
\label{2.4}
&&\delta\varphi=(\partial+\partial\varphi)\epsilon
+\frac{1}{2}\kappa\lambda,\\
&&\delta\lambda=(\epsilon\partial+\frac{1}{2}\partial\epsilon)\lambda
+(\partial+\frac{1}{2}\partial\varphi)\kappa,\nonumber
\end{eqnarray}
which are related to the current fields by the rule
\begin{eqnarray}
\label{2.5}
&&T=\partial^2\varphi-\frac{1}{2}(\partial\varphi)^2+\frac{1}{2}\lambda
\partial\lambda,\\
&&S=\partial\lambda-\frac{1}{2}\lambda\partial\varphi.\nonumber
\end{eqnarray}
we can reduce the order of the derivatives in (\ref{2.1}).
The operator $R$ acts on the fields $\varphi$ and $\lambda$ in the following 
manner:
\begin{eqnarray}
\label{2.6}
&&R_\varphi=(\bar\partial-h\partial)\varphi-\frac{1}{2}\chi\lambda-
\partial h,\\
&&R_\lambda=(\bar\partial-h\partial-\frac{1}{2}\partial h)\lambda-
\frac{1}{2}\chi\partial\varphi-\partial\chi.\nonumber
\end{eqnarray}
We see that the gauge fields $h$ and $\chi$ cannot be expressed in terms of
$\varphi$ and $\lambda$ locally, a situation resembling that of Sect. 2.

The fields $\varphi$ and $\lambda$ form an algebra of free fields in Poisson
bracket:
\begin{eqnarray}
\label{2.7}
&&\{\partial\varphi(x);\partial\varphi(x^\prime
)\}=\delta^\prime(x-x^\prime),\\
&&\{\lambda(x);\lambda(x^\prime)\}=-\delta(x-x^\prime).\nonumber
\end{eqnarray}
Then, with respect to this bracket, (\ref{2.5}) suggests the existence of the
following algebra for the current fields:
\begin{eqnarray}
\label{2.8}
&&-\{T(x);T(x^\prime)\}=\delta^{\prime\prime\prime}(x-x^\prime)+
(T(x)+T(x^\prime))\delta^\prime(x-x^\prime),\\\nonumber
&&-\{T(x);S(x^\prime)\}=(S(x)+\frac{1}{2}S(x^\prime))\delta^\prime
(x-x^\prime),\\
&&-\{S(x);S(x^\prime)\}=\delta^{\prime\prime}(x-x^\prime)-\frac{1}{2}
T(x)\delta(x-x^\prime).\nonumber
\end{eqnarray}
To parametrize the gauge fields in a convenient manner, we must introduce 
a scalar multiplet $(f,\psi)$ without anomalous dimensionality:
\begin{eqnarray}
\label{2.9}
&&\delta f=\epsilon\partial f+\frac{1}{2}\kappa\psi,\nonumber\\
&&\delta\psi=\epsilon\partial\psi+\frac{1}{2}\partial\epsilon\psi
+\frac{1}{2}\kappa\partial f,\\
&&R_f=(\bar\partial-h\partial)f-\frac{1}{2}\chi\psi,\nonumber\\
&&R_\psi=(\bar\partial-h\partial-\frac{1}{2}\partial
 h)\psi-\frac{1}{2}\chi\partial f.\nonumber
\end{eqnarray}
This multiplet is related to the matter fields as follows:
\begin{eqnarray}
\label{2.10}
&&\varphi=\log\partial f+\frac{\psi\partial\psi}{(\partial f)^2},\\
&&\lambda=2\frac{\partial\psi}{\partial
f}-\psi\frac{\partial^2f}{(\partial f)^2}.\nonumber
\end{eqnarray}
There is no simple way in which we can deduce such a complicated relationship
from the condition that the appropriate terms appear in the transformation 
laws. However, the problem can be simplified if superfields are introduced.

Since the superfield formulation of chiral supergravity contains no auxiliary 
fields, the meaning of all previous expressions is not altered when we go over 
to superfields: 
\begin{eqnarray}
\label{2.11}
&&R_U=(\bar\partial-H\partial-\frac{1}{2}DHD-\frac{3}{2}\partial H
)U -\partial^2D H,\nonumber\\
&&\delta U=(D\partial^2+\frac{3}{2}U\partial-\frac{1}{2}DUD+\partial U)E,\\
&&\delta H=(\bar\partial-H\partial-\frac{1}{2}DHD+\partial H)E,\nonumber
\end{eqnarray}
where $U=S+\theta T, H=h+\theta \chi, D=\partial_\theta+\theta\partial$, and
$E=\epsilon+\theta k$ with $\theta$ the anticommuting coordinate.
Then the current superfield $U$ is related to the matter superfield
$\Phi=\varphi+\theta\lambda$ as follows:
\begin{equation}
\label{2.12}
U=D\partial\Phi-\frac{1}{2}D\Phi\partial\Phi,
\end{equation}
and accordingly,
\begin{eqnarray}
\label{2.13}
&&\delta\Phi=\partial E+E\partial\Phi+\frac{1}{2}DED\Phi,\\
&&R_U=D\partial R_\Phi-\frac{1}{2}D\Phi\partial R_\Phi-
\frac{1}{2}\partial\Phi DR_\Phi.\nonumber
\end{eqnarray}
The scalar multiplet $F=f+\theta\psi$, with
\begin{equation}
\label{2.14}
\delta F=E\partial F+\frac{1}{2}DE DF,
\end{equation}
is related to the superfield $\Phi$ by:
\begin{equation}
\label{2.15}
\Phi=\log\partial F+\frac{DF}{\partial F}D\log\partial F,
\end{equation}
while the relationship between corresponding Ward identities is
\begin{equation}
\label{2.16}
R_\Phi=\left(\frac{DF}{(\partial F)^2}\partial D+(\frac{1}{\partial
F}-2\frac{DFD\partial F}{(\partial F)^3})\partial-\frac{D\partial
F}{(\partial F)^2}D\right)R_F.
\end{equation}
The formulae that link the current Ward identities with $R_f$ and $R_\psi$ are
\begin{eqnarray}
\label{2.17}
&&R_T=(\partial^3+2T\partial+\partial T)\left(\frac{R_f}{\partial f}
-R_f\frac{\psi\partial\psi}{(\partial f)^3}+\frac{\psi
R_\psi}{(\partial f)^2}\right)\nonumber\\
&&-(\frac{3}{2}S\partial+\frac{1}{2}\partial S)\left(2\frac{R_\psi}
{\partial f}-\frac{\psi}{\partial f}\partial(\frac{R_f}{\partial
f})-2\frac{\partial\psi R_f}{(\partial f)^2}\right),\\
&&R_S=(\partial^2+\frac{1}{2}T)\left( 2\frac{R_\psi}{\partial f}-
\frac{\psi}{\partial f}\partial(\frac{R_f}{\partial f})-
2\frac{\partial\psi R_f}{(\partial f)^2}\right)\nonumber\\
&&-(\frac{3}{2}S\partial
+\partial S)\left(\frac{R_f}{\partial f}-R_f\frac{\psi\partial\psi}
{(\partial f)^3}+\frac{\psi R_\psi}{(\partial f)^2}\right).\nonumber
\end{eqnarray}
If we now use the Legendre transformations to proceed from the partition 
function to the effective action,
\begin{equation}
\label{2.18}
\Gamma[T,S]=Z[h,\chi]-\int d^2x(hT+\chi S),
\end{equation}
the Ward identity becomes
\begin{equation}
\label{2.19}
(\partial^2D+3U\partial+DUD+2\partial
U)\frac{\delta\Gamma}{\delta U}=0
\end{equation}
or, in components,
\begin{eqnarray}
\label{2.20}
&\left(\partial^3+2T\partial+\partial T\right)\displaystyle{
\delta\Gamma\over\displaystyle\delta T}+\left(\frac{3}{2}S
\partial+\frac{1}{2}\partial S\right)\displaystyle
\frac{\delta\Gamma}{\delta S}=-\bar\partial T,&\\
&\left(\partial^2+\frac{1}{2}T\right)\displaystyle\frac{\delta\Gamma}
{\delta T}+\left(\frac{3}{2}S\partial+\partial S\right)\displaystyle
\frac{\delta\Gamma}{\delta S}=-\bar\partial S,&\nonumber
\end{eqnarray}
i.e., there emerges a supersymmetric Bol operator, which has also been
described in Ref. 12. In this form the covariance of Ward identities
under gauge transformations (\ref{2.2}), which is equivalent to the 
Wess-Zumino conditions for an anomaly, becomes explicit.

Now let us turn to the problem of finding the partition function of the theory:
\begin{eqnarray}
\label{2.21}
&&\delta Z=\int d^2x(T\delta h+S\delta\chi)=\int d^2xd\theta U\delta H\\
&&=-\int d^2xd\theta E(\bar\partial-H\partial-\frac{1}{2}DHD+
\frac{1}{2}\partial H)U.\nonumber
\end{eqnarray}
We see that the integrand is the Ward identity $R_U$ without anomalous term.
In the chiral scheme, i.e., a regularization scheme that conserves half the 
reparametrization symmetry and the Weyl symmetry as well as their 
superpartners, the fields $\varphi$ and $\lambda$ are related by (\ref{2.10}), 
and the corresponding Ward identities $R_\varphi$ and $R_\lambda$ vanish.
On the other hand, in a regularization scheme that preserves supercoordinate
symmetry the group parameters $f$ and $\psi$ vanish and the fields $\varphi$ 
and $\lambda$ are unconstrained, but Ward identities $R_\varphi$ and 
$R_\lambda$ still play an important role.
Multiplaying $R_U$ by $E$ and  integrating by parts, we obtain
\begin{eqnarray}
\label{2.22}
&&\int d^2xd\theta R_UE=\int E(D\partial-\frac{1}{2}D\Phi\partial-
\frac{1}{2}\partial\Phi D)R_\Phi\nonumber\\
&&=\int d^2xd\theta R_\Phi(D\partial E+\frac{1}{2}\partial
(ED\Phi)+\frac{1}{2}D(E\partial\Phi))\\
&&=\int d^2xd\theta R_\Phi D\delta\Phi.\nonumber
\end{eqnarray}
If we "err" twice, i.e., take (\ref{2.22}) for $\delta Z$ rather than
(\ref{2.21}) and ignore  the relationship between $H$ and $\Phi$,
expressed by the fact that $R_\Phi$ is zero, we can integrate this 
variational equation and arrive at the following expression for $Z[H]$:
\begin{equation}
\label{2.23}
Z[H]=-\frac{1}{2}\int d^2xd\theta(\bar\partial\Phi-H\partial
\Phi-2\partial H)D\Phi.
\end{equation}
The fact that (\ref{2.17}) reproduces the anomaly correctly can easily be 
verified. Thus, assuming that the superfield 
$\Phi$ is independent, we can reproduce the anomaly by directly adding the 
appropriate term to the action.

This conclusion agrees with our ideas about the "transfer" of the anomaly
from one regularization scheme to another. A detailed description of the
process in which a conformal anomaly is transformed into a gravitational 
anomaly in the two-dimensional gravitation theory can be found in \cite{rrd}.
\setcounter{equation}{0}

\section{$W_3$-gravity}

The difference between the theory of $W_3$-gravity and above cases in that
the chiral formulation of this theory is not only more convenient but is also
the only one amenable to quantum analisis. The nonchiral version formulated 
in \cite{ssn1}, contains an infinite number of derivatives of matter fields and
is too complicated even at the classical level.

The Ward identities of chiral $W_3$-gravity are
\begin{eqnarray}
\label{3.1}
R_T&=&(\bar\partial-h\partial-2\partial h)T-(2b\partial+
3\partial b)W-\partial^3h,\nonumber\\
R_W&=&(\bar\partial-h\partial-3\partial h)W+(2b\partial^3+
9\partial B\partial^2\\&+&15\partial^2b\partial+10\partial^3b
+16bT\partial+16\partial bT)T-\partial^5h.\nonumber
\end{eqnarray}
Here $b$ denotes the single nonvanishing component of a third rank symmetric 
tensor - the gauge field of $W$-gravity, the partner of the metric in the 
multiplet - and $W$ denotes the corresponding spin-3 current, the partner of
the energy-momentum tensor. The chiral general-coordinate and $W$-transformations 
have the form
\begin{eqnarray}
\label{3.2}
&&\delta T=(\partial^3+2T\partial+\partial T)\epsilon+
3W\partial\lambda+2\partial W\lambda,\nonumber\\
&&\delta W=\partial W\epsilon+3W\partial\epsilon+(\partial^5+10T\partial
^3+15\partial T\partial^2\nonumber\\
&&+9\partial^2T\partial+2\partial^3T
+16T^2\partial+16T\partial T)\lambda,\\
&&\delta h=(\bar\partial-h\partial+\partial h)\epsilon+2\lambda
\partial^3b\nonumber\\
&&-3\partial\lambda\partial^2b+3\partial^2\lambda\partial
b-2\partial^3b+16T(\lambda\partial b-b\partial\lambda),\nonumber\\
&&\delta b=\epsilon\partial b-2\partial\epsilon
b+(\bar\partial-h\partial+2\partial h)\lambda.\nonumber
\end{eqnarray}
The quantities $R_T$ and $R_W$ are covariant under the transformations 
(\ref{3.2}), i.e.,
\begin{eqnarray}
\label{3.3}
\delta R_T&=&(\epsilon\partial+2\partial\epsilon)R_T+
(2\lambda\partial+3\partial\lambda)R_W,\nonumber\\
\delta R_W&=&(\epsilon\partial+3\partial\epsilon)R_W+
(2\lambda\partial^3
+9\partial\lambda\partial^2+15\partial^2\lambda\partial\\
&+&10\partial^3\lambda+32T\partial\lambda+16T\lambda\partial+
16\lambda\partial T)R_T.\nonumber
\end{eqnarray}
Thus, taking $R$ as the differential operator, we conclude that (\ref{3.3}) 
yields a universal relation
\begin{equation}
\label{3.4}
[\delta,R]=0
\end{equation}
on the current fields $T$ and $W$.

The transformations (\ref{3.2}) of an arbitrary quantity $A$ are generated
by the currents $T$ and $W$ via Poisson brackets:
\begin{equation}
\label{3.5}
\delta A=\int d^2x\left(\epsilon(x)\{T(x);A\}+\lambda(x)
\{W(x);A\}\right).
\end{equation}
In terms of these brackets, the transformations (\ref{3.2}) themselves become
\begin{eqnarray}
\label{3.6}
&&-\{T(x);T(x^\prime)\}=\delta^{\prime\prime\prime}(x-x^\prime)+
(T(x)+T(x^\prime))\delta^\prime(x-x^\prime),\nonumber\\
&&-\{T(x);W(x^\prime)\}=(W(x)+2W(x^\prime))\delta^\prime(x-x^\prime),
\nonumber\\
&&-\{W(x);T(x^\prime)\}=(2W(x)+W(x^\prime))\delta^\prime(x-x^\prime),\\
&&-\{W(x);W(x^\prime)\}=\delta^V+5(T(x)+T(x^\prime))\delta^{\prime
\prime\prime}(x-x^\prime)\nonumber\\&&+8(T^2(x)+T^2(x^\prime))\delta^
\prime(x-x^\prime)-3(T^{\prime\prime}(x)+T^{\prime\prime}(x^\prime))
\delta^\prime(x-x^\prime).\nonumber
\end{eqnarray}
This algebra can be reproduced by expressing the current fields in terms of 
the matter fields $\varphi$ and $\psi$, which obey the algebra of free fields'
\begin{eqnarray}
\label{3.7}
\{\varphi^\prime(x);\varphi^\prime(x^\prime)\}&=&\delta^\prime
(x-x^\prime),\nonumber\\
\{\varphi^\prime(x);\psi^\prime(x^\prime)\}&=&0,\\
\{\psi^\prime(x);\psi^\prime(x^\prime)\}&=&-\delta^\prime
(x-x^\prime),\nonumber
\end{eqnarray}
if we define $T(x)$ and $W(x)$ in the following manner:
\begin{eqnarray}
\label{3.8}
T(x)=\partial^2\varphi-\frac{1}{2}(\partial\varphi)^2+\frac{1}{2}
(\partial\psi)^2,\nonumber\\
W(x)=\partial^3\psi-3\partial\varphi\partial^2\psi-\partial^2\varphi
\partial\psi\\
+2(\partial\varphi)^2\partial\psi+\frac{2}{3}(\partial\psi)^3.\nonumber
\end{eqnarray}

This corresponds to the following transformation law for the matter fields:
\begin{eqnarray}
\label{3.9}
&&\delta\varphi=\partial\epsilon+\partial\varphi\epsilon-4\lambda\partial
\varphi\partial\psi+2\lambda\partial^2\psi-\partial\lambda\partial\psi,\\
&&\delta\psi=\epsilon\partial\psi+\partial^2\lambda+3\partial\lambda
\partial\varphi+2\lambda(\partial^2\varphi+(\partial\varphi)^2+
(\partial\psi)^2).\nonumber
\end{eqnarray}

The anomalous equation of motion of the matter multiplet have the form
\begin{eqnarray}
\label{3.10}
&&R_\varphi=\bar\partial\varphi-h\partial\varphi+\partial
b\partial\psi-2b(\partial^2\psi-2\partial\varphi\partial\psi)
-\partial h,\\
&&R_\psi=\bar\partial\psi-h\partial\psi-3\partial b\partial\varphi
-2b(\partial^2\varphi+(\partial\varphi)^2+(\partial\psi)^2)-
\partial^2b.\nonumber
\end{eqnarray}
With respect to $\epsilon$- and $\lambda$-diffeomorphysms, these relationships
are also covariant
\begin{eqnarray}
\label{3.11}
&\delta R_\varphi=(\epsilon\partial-4\lambda\partial\psi\partial)
R_\varphi+(2\lambda\partial^2-\partial\lambda-4\lambda\partial\varphi
\partial)R_\psi,\\
&\delta R_\psi=(\epsilon\partial+4\lambda\partial\psi)R_\psi+
(3\partial\lambda\partial+2\lambda\partial^2+4\lambda\partial
\varphi\partial)R_\varphi&.\nonumber
\end{eqnarray}
This also estabilishes the validity of Eq. (\ref{3.4}) when acts on matter
multiplet.

The Ward identiites (\ref{3.1}) can easily be transformed into 
\begin{eqnarray}
\label{3.12}
&&R_T=(\partial^2-\partial\varphi)R_\varphi+\partial\psi\partial
 R_\psi\nonumber,\\
&&R_W=(4\partial\varphi\partial\partial\psi\partial-\partial\psi
\partial^2-3\partial^2\psi\partial)R_\varphi\\
&&+(\partial^3-3\partial\varphi\partial^2+2(\partial\varphi)^2
\partial+2(\partial\psi)^2\partial)R_\psi.\nonumber
\end{eqnarray}
The $\epsilon$- and $\lambda$-transformations constitute a closed algebra on
the multiplets of currents $\{T,W\}$, matter fields $\{\varphi,\psi\}$ and
gauge fields $\{h,b\}$:
\begin{eqnarray}
\label{3.13}
{[}\delta(\epsilon_1),\delta(\epsilon_2){]}&=&\delta(\epsilon_3=
\epsilon_2\partial\epsilon_1-\epsilon_1\partial\epsilon_2),\nonumber\\
{[}\delta(\epsilon_1),\delta(\lambda_2){]}&=&\delta(\lambda_3=
2\lambda_2\partial\epsilon_1-\epsilon_2\partial\lambda_1)\\
{[}\delta(\lambda_1);\delta(\lambda_2){]}&=&\delta(\epsilon_3=
16T(\lambda_2\partial\lambda_1-\lambda_1\partial\lambda_2),\nonumber\\
&+&2\lambda_2\partial^3\lambda_1-3\partial\lambda_2\partial^2\lambda
_1+3\partial^2\lambda_2\partial\lambda_1-2\partial^3\lambda_2\lambda
_1).\nonumber
\end{eqnarray}
The partition function of the theory is calculated in the same way as in 
supergravitation theory: by multiplaying Eqs. (\ref{3.12}) respectively,
by $\epsilon$ and $\lambda$, we obtain
\begin{equation}
\label{3.14}
\int d^2x(\epsilon R_T+\lambda R_W)=\int d^2x(R_\varphi\delta\partial
\varphi-R_\psi\delta\partial\psi).
\end{equation}

If the Ward identities (\ref{3.1}) were to have no anomalous terms, the 
left-hand side of Eq. (\ref{3.14}) would be the variation of the partition 
function with opposite sign. If, in addition, the fields $\varphi$ and
$\psi$ were to be free and Eqs. (\ref{3.10}) were not link them with 
gauge fields $h$ and $b$, the right-hand side of Eq. (\ref{3.14}) would be 
the total variation of the following expression:
¢ëà ¦¥­¨ï:
\begin{eqnarray}
\label{3.15}
Z[h,b]=\int d^2x\left(\frac{1}{2}\partial\varphi(\bar\partial
\varphi-h\partial\varphi)-\frac{1}{2}\partial\psi(\bar\partial\psi-
h\partial\psi)-\partial h\partial\varphi+\right.\\\nonumber
\left.+\partial^2b\partial\psi+b(2(\partial\varphi)^2\partial\psi-
\partial^2\varphi\partial\psi-3\partial\varphi\partial^2\psi
+\frac{2}{3}(\partial\psi)^3)\right)\\
\quad =\int d^2x[\frac{1}{2}(\bar\partial\varphi\partial\varphi-
\bar\partial\psi\partial\psi)+hT+bW].\nonumber
\end{eqnarray}
Equation (\ref{3.15}) is the action of the matter fields interacting with
two-dimensional chiral gravity and $W_3$-gravity. The variation of (\ref{3.15})
with respect to (\ref{3.2}) and (\ref{3.9}) is
\begin{equation}
\label{3.16}
\delta Z=\int d^2x\left(h\partial^3\epsilon+b\partial^5\lambda+
16\lambda(T^2\partial b+bT\partial T)\right).
\end{equation}
This expression differs from that for the quantum anomaly of the minimal type
by the presence of terms quadratic in $T$. This discrepancy is due to the 
differences in defining the transformation law for the field $h$ under 
$\lambda$-diffeomorphysms. The transformation law (\ref{3.2}) is motivated
by the closure of the algebra (\ref{3.13}) on the fields $h$ and $b$ and by
the Wess-Zumino self-consistency condition, with Eqs. (\ref{3.3}) being valid.

The following line of reasoning can motivate another definition of 
$\delta_\lambda h$ : if we perform the Legendre transformation
\begin{equation}
\label{3.17}
Z[h,b]=\Gamma[T,W]+\int d^2x(hT+bW),
\end{equation}
the expressions for $R_T$ and $R_W$ acquire the Bol operators $L_3$ and $L_5$:
\begin{eqnarray}
\label{3.18}
R_T=\bar\partial T+\left(\partial^3+2T\partial+\partial
T\right)\frac{\delta\Gamma}{\delta T}+\left( 3W\partial+2
\partial W\right)\frac{\delta \Gamma}{\delta W},\\\nonumber
R_W=\bar\partial W+\left( 3W\partial+\partial W\right)\frac
{\delta\Gamma}{\delta T}+\left(\partial^5+10T\partial^3+15
\partial T\partial^2+\right.\\
\left.+(9\partial^2T+16T^2)\partial+(2\partial^3T+16T\partial
T)\right)\frac{\delta\Gamma}{\delta W}.\nonumber
\end{eqnarray}
Thus,
\begin{eqnarray}
\label{3.19}
\delta Z&=&\int d^2x(T\delta h+W\delta b)\\
&=&-\int d^2x\left(\epsilon\partial^3 h+\lambda\partial^5
b-16\lambda(T^2+bT\partial T)\right).\nonumber
\end{eqnarray}
To obtain an anomaly of the minimum type it appears reasonable to define
$\delta_\lambda h$ with $\delta_{extra}h=-8T(\lambda\partial
b-b\partial\lambda)$.

The anomaly can be completely removed from the transformation laws and 
the Ward identities if we transform to variables $(f,g)$, which form a scalar
multiplet:
\begin{eqnarray}
\label{3.20}
&&\varphi=\log\partial f+\frac{1}{2}\log(1+\not\!\partial^2g),\\
&&\psi=\gamma^{-1}\log(1+\not\!\partial^2g),\nonumber
\end{eqnarray}
where $\not\!\partial\equiv\frac{1}{\partial f}\partial$ and
$\gamma^2=-12$

The transformation law for the fields $f$ and $g$ is
\begin{eqnarray}
\label{3.21}
&&\delta f=\epsilon\partial f-\gamma(\lambda\partial^2f+
\frac{1}{2}\partial\lambda\partial f +\frac{2}{3}\lambda\partial
f\partial\log(1+\not\!\partial^2g)),
\\
&&\delta g=\epsilon\partial g-\frac{1}{2}\gamma\partial\lambda
\partial g-
\gamma\lambda[(\partial f)^2-\partial^2g
+\frac{2}{3}\partial g\partial\log(1+\not\!
\partial^2g)+2\partial g\frac{\partial^2f}{\partial f}].\nonumber
\end{eqnarray}
Accordingly, the Ward identiites are
\begin{eqnarray}
\label{3.22}
&&R_f=\bar\partial f-h\partial f+\frac{\gamma}{2}\partial b
\partial f+\gamma b(\partial^2f+\frac{2}{3}\partial
f\partial\log(1+\not\!\partial^2g))
\\
&&R_g=\bar\partial g-h\partial g+\frac{\gamma}{2}\partial
b\partial f
+\gamma b[2\partial g\frac{\partial^2f}{\partial f}
+\frac{2}{3}\partial f\partial\log(1+\not\!\partial^2g)-
(\partial f)^2-\partial^2g].\nonumber
\end{eqnarray}
Setting $R_f$ and $R_g$ to zero, we can express the gauge multiplet in terms
of $f$ and $g$:
\begin{eqnarray}
\label{3.23}
&&b=\gamma^{-1}\frac{(\bar\partial g-\frac{\bar\partial
f}{\partial f}\partial g)}{(\partial f)^2(1+\not\!\partial^2g)},\\
&&h=\frac{\bar\partial f}{\partial f}+\frac{\gamma}{2}\partial b+
\gamma b(\frac{\partial^2f}{\partial f}+\frac{2}{3}\partial\log
(1+\not\!\partial^2g)).
\end{eqnarray}
These formulae coincide, up to renormalization of $\gamma$ and $\lambda$, with
the solution found by Ooguri et. al., \cite{ossn}, who interpreted
$W_3$-gravity as a constrained Wess-Zumino $SL(3,R)$-theory. Plugging
(\ref{3.20}) and (\ref{3.21}) into (\ref{3.15}), we reproduce the chiral 
action of Ooguri et. al., \cite{ossn}, which must be interpreted as the
effective action induced by quantum fluctuations of the matter fields $\varphi$
and $\psi$, which interact with the multiplet of chiral $W_3$-gravity via 
(\ref{3.15}). The anomalous dependence of this action on the "coordinate"
functions $f$ and $g$ is due to the $W$-gravity anomaly. Continuing the
analogy with the cases of two-dimensional gravity and supergravity, we 
can assume that this action is a result of choosing a regularization scheme
that conversed the Weyl and $W$-Weyl symmetries , as well as half the 
coordinate symmetries of the covariant action that describes the interaction 
of matter fields and ordinary gravity and $W$-gravity.

\setcounter{equation}{0}
\section{Conclusion}

By generalizing some of the laws governing ordinary gravitation we were able to
find the effective action of chiral $W_3$-gravity. It would be more interesting, 
however, to continue the analogy and find the $W$-analog of the Liouville 
action; namely the covariant action that describes the interaction of matter 
fields with a gravitational and $W$-gravitational background.

To understand the nature of the symmetry properties of $W$-gravity theory it is 
advisable to first turn to classical theory.

As a classical gauge theory, $W$-gravity was first examined by Hull
\cite{h}. The nonchiral formulation of this theory was later
performed by Schoutens et. al. \cite{ssn1}. They started with the action
\begin{equation}
\label{4.1}
S=\frac{1}{2}\int d^2x\partial_+\varphi\partial_-\varphi.
\end{equation}
With respect to the ordinary diffeomorphysms 
$\delta\varphi=\epsilon^\alpha\partial_\alpha\varphi$
the variation of (\ref{4.1})
\begin{equation}
\label{4.2}
\delta S=\int d^2x[\partial_\alpha(\epsilon^\alpha\partial_+
\varphi\partial_-\varphi)+\partial_+\epsilon^-(\partial_-\varphi)^2
+\partial_\epsilon^+(\partial_+\varphi)^2,
\end{equation}
vanishes if $\partial_+\epsilon^-=\partial_-\epsilon^+=0$.
To generalize this symmetry to a local one, we must, in accordance with
Noether's theorem, add to (\ref{4.1}) the currents $t_{++}=
\frac{1}{2}(\partial_+\varphi)^2$ and $t_{--}=\frac{1}{2}
(\partial_-\varphi)^2$ multiplied by the corresponding gauge fields $kh_{++}$ 
and $kh _{--}$. Here $k$ is an expansion parameter, which is set to unity in
the final result. After an infinite number of steps the action can be summed 
as a geometric progression to produce
\begin{equation}
\label{4.3}
S=\frac{1}{2}\int
d^2x\frac{(\partial_+\varphi-kh^{--}\partial_-\varphi)
(\partial_-\varphi-kh^{++}\partial_+\varphi}{1-k^2h^{++}h^{--}}.
\end{equation}
Schoutens et. al. \cite{ssn1} then stated that the action (\ref{4.1})
for $n$ real scalar fields is invariant under holomorphic 
$W$-diffeomorphysms. Indeed, under the transformations
\begin{equation}
\label{4.4}
\delta\varphi^i=d^{ijk}(\lambda^{++}\partial_+\varphi^j\partial_+
\varphi^k+\lambda^{--}\partial_-\varphi^j\partial_-\varphi^k)
\end{equation}
the action transforms as
\begin{equation}
\label{4.5}
\delta S=\frac{1}{3}\int d^2xd_{ijk}(\partial_+\varphi^i
\partial_+\varphi^j\partial_+\varphi^k\partial_-\lambda^{++}
+\partial_-\varphi^i\partial_-\varphi^j\partial_-\varphi^k
\partial_+\lambda^{--}).
\end{equation}
The fact that the algebra of holomorphic $\epsilon$- and $\lambda$-
diffeomorphysms is closed imposes the following constraint on the symmetric 
constants $d$ see \cite{h}:
\begin{equation}
\label{4.6}
d^{k(ij}d^{l)mk}=\delta^{(ij}\delta^{l)m}.
\end{equation}
The action (\ref{4.1}) can be made invariant under local $\epsilon$- and 
$\lambda$-transformations via the Noether procedure by introducing the
appropriate gauge fields $h^{++}$, $h^{--}$, $b^{+++}$ and $b^{---}$.
Unfortunately, in the given case the invariant action can be summed only
by using auxiliary fields $F^i_{\pm}$:
\begin{eqnarray}
\label{4.7}
S=\int d^2xe[\partial_+\varphi^i\partial_-\varphi^i+F^i_+F^i_-
+F^i_+(\partial_-\varphi^i-\frac{1}{3}d_{ijk}b^{+++}F^j_+F^k_+)+\\
F^i_-(\partial_+\varphi^i-\frac{1}{3}d_{ijk}b^{---}F^j_-F^k_-)].
\nonumber
\end{eqnarray}
After the auxiliary fields are eliminated, they replaced by "nested" covariant
derivatives,
\begin{equation}
\label{4.8}
F^i_{\pm}\rightarrow\hat\partial\varphi^i_{\pm}=
e^\alpha_{\pm}\partial_\alpha\varphi-b^{\mp\mp\mp}d_{ijk}
\hat\partial_{\pm}\varphi^j\hat\partial_{\pm}\varphi^k,
\end{equation}
and the action becomes infinitely nonlinear. To avoid difficulties and operate
from the start in a covariant setting, we must introduce more gauge fields 
than required by the Noether approach. Specifically, in the case of pure 
gravitation, we must introduce the full tensor $h_{\alpha\beta}$ instead of 
the two components $h{++}$ ¨ $h{--}$. Then the Noether procedure terminates 
after the first step, and the invariant action has the form
$$S=\int d^2xh^{\alpha\beta}\partial_\alpha\varphi\partial_
\beta\varphi.$$

Here we expect a new symmetry to appear, a symmetry that would balance
the superfluous degree of freedom related to the $h{+-}$-component of
the metric. By requiring that the energy-momentum tensor of the theory be
traceless,
\begin{equation}
\label{4.9}
T^\alpha_\alpha=T_{\alpha\beta}h^{\alpha\beta}=0,\quad\quad
T_{\alpha\beta}\equiv\frac{\delta S}{\delta h^{\alpha\beta}},
\end{equation}
we require that the theory be Weyl-invariant, so that the equation 
$h^{\alpha\beta}\frac{\delta S}{\delta h^{\alpha\beta}}=0$ has a functional of 
the type $S=S(\sqrt hh^{\alpha\beta})$ as a solution, which is equivalent to 
invariance under $h^{\alpha\beta}\rightarrow e^\sigma h^{\alpha\beta}$. In 
this way the final expression for the invariant action is
\begin{equation}
\label{4.10}
S=\int\sqrt hd^2xh^{\alpha\beta}\partial_\alpha\varphi\partial_
\beta\varphi.
\end{equation}
In the case of $W_3$-gravity we propose introducing the $h^{+-}, b^{++-}$
and $b^{--+}$ components of the gauge fields, in order to produce the total
tensors $h^{\alpha\beta}, b^{\alpha\beta\gamma}$. The Noether procedure 
terminates after the first step, and the invariant action has the form 
\begin{equation}
\label{4.11}
S=\int d^2x(h^{\alpha\beta}t_{\alpha\beta}+
b^{\alpha\beta\gamma}\omega_{\alpha\beta\gamma}),
\end{equation}
where
$t_{\alpha\beta}=\frac{1}{2}(\partial_\alpha\varphi\partial_
\beta\varphi-\partial_\alpha\psi\partial_\beta\varphi)$ and
$$\omega_{\alpha\beta\gamma}=\frac{2}{3}(\partial_\alpha\varphi
\partial_\beta\varphi\partial_\gamma\psi+\partial_\alpha\varphi
\partial_\beta\psi\partial_\gamma\varphi+\partial_\alpha\psi
\partial_\beta\varphi\partial_\gamma\varphi+\partial_\alpha\psi
\partial_\beta\psi\partial_\gamma\psi)$$
The action (\ref{4.11}) is assumed to be invariant under the $W$-
diffeomorphysms 
\begin{eqnarray}
\label{4.12}
&&\delta_\lambda\varphi=-4\lambda^{\alpha\beta}\partial_\alpha
\varphi\partial_\beta\psi,\\
&&\delta_\lambda\psi=2\lambda^{\alpha\beta}(\partial_\alpha\varphi
\partial_\beta\varphi+\partial_\alpha\psi\partial_\beta\psi),\nonumber
\end{eqnarray}
defined with a traceless parameter $\lambda$, i.e., $\lambda^{\alpha
\beta}h_{\alpha\beta}=0$. The variation of (\ref{4.11}) under the
transformations (\ref{4.12}) can be written in the form
\begin{eqnarray}
\label{4.13}
\delta S=\int d^2x
\left[\delta_\lambda b^{\alpha\beta\gamma}\omega_
      {\alpha\beta\gamma}-\omega_{\alpha\beta\gamma}(h^{\alpha\mu}
      \nabla_\mu\lambda^{\beta\gamma}+h^{\beta\mu}\nabla_\mu\lambda
      ^{\alpha\gamma}\right.
+h^{\gamma\mu}\nabla_
\mu\lambda^{\alpha\beta})\nonumber\\
-(h^{\alpha\mu}\lambda^{\beta\gamma}+h^{\beta\gamma}\lambda
^{\alpha\mu})\nabla_\mu\omega_{\alpha\beta\gamma}+\delta_\lambda
h^{\alpha\beta}t_{\alpha\beta}+16b^{\alpha\beta\gamma}\lambda^{\mu\nu}\\
  \times (2t_{\gamma\nu}\nabla_\mu t_{\alpha\beta}-t{\beta\gamma}
\nabla_\alpha t_{\mu\nu})
+16b^{\alpha\beta\gamma}\nabla_\alpha\lambda^{\mu\nu}(2t_{\beta
\mu}t_{\gamma\nu}-t_{\beta\gamma}t_{\mu\nu})\left.\right].
\nonumber
\end{eqnarray}
Defining the $\lambda$-variations of the gauge fields in such a way 
that the coefficients of the currents $t_{\alpha\beta}$ and
$\omega_{\alpha\beta\gamma}$ vanish, we ensure that the action 
(\ref{4.11}) is invariant.
The transformations (\ref{4.12}) represent a specific realization 
of the constants $d^{ijk}$ in (\ref{4.4}) for the case of two fields.
Such a restriction is not accidental, the point being that when there 
are three or more fields, the condition that the covariant algebra of the
$W$-transformations
\begin{equation}
\label{4.14}
\delta\varphi^i=d^{ijk}\lambda^{\alpha\beta}\partial_\alpha
\varphi^j\partial_\beta\varphi^k)
\end{equation}
be closed imposes additional constrants on the $d^{ijk}$, restrictions
which together with the conditions (\ref{4.6}) have only a vanishing
solution. In addition, the algebra becomes closed on the gauge fields 
only if the equations of motion are included, as happens in the simpler 
case of two-dimwnsional supergravitation (\cite{dz}).

Thus, to guarantee invariance under ordinary and $\lambda$-
diffeomorphysms, we introduced seven gauge fields. Now we would
like to impose constraints on the theory in such a way so as to obtain
a three-parameter symmetry group that "balances" the three superfluous
degrees of freedom.

By imposing the conditions that the energy-momentum tensor and 
$W$-current be traceless,
\begin{equation}
\label{4.15}
h^{\alpha\beta}\frac{\delta S}{\delta h^{\alpha\beta}}=0\quad
h^{\alpha\beta}\frac{\delta S}{\delta b^{\alpha\beta\gamma}}
\end{equation}
conditions with a solution of the form $S=S(\hat h^{\alpha\beta},\hat
b^{\alpha\beta\gamma})$, with the quantities $\hat h^
{\alpha\beta}\equiv\sqrt hh^{\alpha\beta}$ and $\hat b^{\alpha
\beta\gamma}\equiv b^{\alpha\beta\gamma}-\frac{1}{4}h_{\mu\nu}
(h^{\alpha\beta}b^{\gamma\mu\nu}+h^{\beta\gamma}b^{\alpha\mu\nu}
+h^{\alpha\gamma}b^{\beta\mu\nu})$ invariant under Weyl and $W$-Weyl
transformations, respectively, i.e.,
\begin{eqnarray}
\label{4.16}
&&Weyl:\quad\quad\quad h^{\alpha\beta}\rightarrow e^
\sigma h^{\alpha\beta},\\
&&W-Weyl:\quad b^{\alpha\beta\gamma}\rightarrow
b^{\alpha\beta\gamma}+(\zeta^\alpha h^{\beta\gamma}+\zeta^\beta h
^{\alpha\gamma}+\zeta^\gamma h^{\alpha\beta})\nonumber
\end{eqnarray}
we reduce the number of degrees of freedom. However, comparison of the
variations $\delta S(\hat h,\hat b)$ with (\ref{4.13}) shows that the
symmetry of the action under $\lambda$-diffeomorphysms is 
incompatible with Weyl invariance, since if the variation
$\delta_\lambda t_{\alpha\beta}$ traceless, the variation
$\delta_\lambda\omega_{\alpha\beta\gamma}$ is not, with the 
result that the latter cannot be made equal to the traceless
variation $\delta \hat h$.
Thus, Weyl invariance is incompatible with $W$-symmetry even at
classical level, and to reduce the number of degrees of freedom 
we must replace the requirements that the energy-momentum tensor 
be traceless by a different one.

Schoutens et. al. \cite{ssn3} proposed a covariant formulation 
of $W$-gravity. With each annihilation operator of the classical 
$W_3$-algebra they assosiated a gauge field and a local parameter, 
and with each creation operator they assosiated a field in the
adjoint representation. Then they required that all corresponding
curvatures vanish. The resulting theory has a finite number of 
degrees of freedom and, in addition, to being coordinate- and $W$-
diffeomorphysm invariant, it is locally Weyl- and Lorentz- and $W$-
Weyl and $W$-Lorentz-invariant.

It would also be interesting to study this theory as a constrained
Hamiltonian system.

Many thanks go to R.L. Mkrtchyan and O.M. Khudaverdyan for fruitful 
and stimulating discussions and especially to A.G. Sedrakyan for a
through review of manuscript and for the critical remarks. The 
present work has in part been supported by grants 211-5291 YPI
(German Bundesministerium fur Forschung und Technologie) and
INTAS-2058.


\begin{thebibliography}{10}
\bibitem{z} A.B. Zamolodchikov, Teoret. Mat. Fiz. 65, (1985), 347.
\bibitem{GD} I.M. Gel'fand and L.A.Dikij Preprint IFM of the USSR Academy 
of Science, Moskow (1978); V.G. Drinfel'd and V.V.Sokolov, Current Problems
of Mathematics [in Russian] Vol.24, VINITI, Moskow (1984).
\bibitem{h} C.M.Hull Phys.Lett. {\bf B240}, (1990), 110.
\bibitem{ssn1} K.Schoutens, A.Sevrin and P.van Nieuvenhuizen,
              Phys.Lett. {\bf B243}, (1990), 248.
\bibitem{ssn2} K.Schoutens, A.Sevrin and P.van Nieuvenhuizen,
              Nucl.Phys. {\bf B364}, (1991), 584; Nucl.Phys.
              {\bf B371}, (1992),315.
\bibitem{ossn} H.Ooguri, K.Schoutens, A.Sevrin and P.van Nieuvenhuizen
               Comm.Math.Phys. {\bf 145}, (1992), 515.
\bibitem{pol} A.M.Polyakov, Mod.Phys.Lett.{\bf A2}, (1987), 893.
\bibitem{kpz} V.G.Knizhnik, A.M.Polyakov, A.B.Zamolodchikov,
              Mod.Phys.Lett.{\bf A3} (1988) 819.
\bibitem{zu} R.Zucchini, Phys.Lett. {\bf B260} (1991) 296.
\bibitem{g} F.Gieres, Conformally covariant operators on Riemann
            Surfaces (with applications to conformal field theory
            and integrable models), CERN  preprint 366/1991.
\bibitem{rrd} D.R.Karakhanyan, R.P.Manvelyan, R.L.Mkrtchyan
              Phys.Lett. {\bf B329}, (1994), 185.
\bibitem{fh} J.M.Figueroa-O'Farrill, S.Stanciu, E.Ramos, Phys.Lett.
             {\bf B297} (1992), 289.
             C.M.Hull, Phys.Lett. {\bf B269}, (1991), 257.
\bibitem{m} Y.Matsuo, Phys.Lett. {\bf B227}, (1991), 117.
\bibitem{pz} A.M.Polyakov, and A.B.Zamolodchikov,
              Mod.Phys.Lett.{\bf A3} (1988) 819.
\bibitem{k} D.R.Karakhanyan, Phys.Lett. {\bf B365}, (1996), 56.
\bibitem{ssn3} K.Schoutens, A.Sevrin and P.van Nieuvenhuizen,
             Nucl.Phys. {\bf B349}, (1991), 791.
\bibitem{dz} S.Deser, B.Zumino, Phys.Lett. {\bf B65}, (1976), 369.


\end{thebibliography}
\end {document}